\documentclass[a4paper,twocolumn,aps,pra,showpacs]{revtex4}
\usepackage[latin9]{inputenc}
\usepackage{float}
\usepackage{amsmath}
\usepackage{amssymb}
\usepackage{graphicx}
\usepackage{esint}

\makeatletter

\@ifundefined{textcolor}{}
{%
 \definecolor{BLACK}{gray}{0}
 \definecolor{WHITE}{gray}{1}
 \definecolor{RED}{rgb}{1,0,0}
 \definecolor{GREEN}{rgb}{0,1,0}
 \definecolor{BLUE}{rgb}{0,0,1}
 \definecolor{CYAN}{cmyk}{1,0,0,0}
 \definecolor{MAGENTA}{cmyk}{0,1,0,0}
 \definecolor{YELLOW}{cmyk}{0,0,1,0}
}

\usepackage{mathrsfs}\usepackage{subfigure}\usepackage{multirow}\usepackage[title,titletoc,header]{appendix}

\makeatother
\begin{document}

 \draft
% \title{Josephson's relation for multiple component bosons and quantum hydrodynamic for spin orbit coupled Bose-Einstein condensates }
 \title{Superfluid density, Josephson relation and pairing fluctuations in a  multi-component fermion superfluid }
 \author{Yi-Cai Zhang}
 \address{Department of Physics, School of Physics and Materials Science, Guangzhou University, Guangzhou
510006, China,
% $^{2}$Department of Physics and Center of Theoretical and Computational Physics, The University  of Hong Kong, Hong Kong, China
  }
\date{\today}

\begin{abstract}
In this work, a Josephson relation is generalized to a  multi-component fermion superfluid. Superfluid density is expressed through a two-particle Green function for pairing states.  When the system has only one gapless collective excitation mode, the Josephson relation is simplified, which is given in terms of  the order parameters and the trace of two-particle normal Green function. In addition, it is found that the two-particle Green function is directly related to the pairing fluctuations of order parameters. Further more,
in the presence of inversion symmetry, the superfluid density is given in  terms of the pairing fluctuation matrix.
  The results of the superfluid density in Haldane model show that the generalized Josephson relation can be also applied to a multi-band fermion superfluid in lattice.
\end{abstract}

\maketitle
\section{Introduction }
The superfluid density $\rho_s$ and order parameter $n_0$ in superfluid liquid Helium-4 are two closely related  \cite{Tisza,BOGOLIUBOV1947}, but different concepts \cite{Penrose,Pollock}. However, they can be connected to each other through a  Josephson relation \cite{Bogoliubov61,Josephson1966,Hohenberg1965,Bogoliubov}, i.e.,
 \begin{align}
\rho_s=-{\rm lim_{q\rightarrow0}}\frac{n_0m}{q^2 G(\textbf{q},0)},\label{bas0}
\end{align}
 where $\rho_s $ is superfluid density (particle number per unit volume), $n_0$ is order parameter (condensate density) in liquid Helium-4. $G(\textbf{q},0)$ is normal single-particle Green function at zero frequency, $m$ is particle mass.
 The above equation indicates that the Green function diverges as wave vector $q\rightarrow0$.
  Such a divergence of $1/q^2$ in Green function is quite a universal  phenomenon which can occur in many systems, e.g., superfluid Helium-4, superconductor and ferromagnetism \cite{Forster}.
%In spontaneously broken symmetry, the correlations function show a singularity in long wave limit.
 % In fact, this can be viewed an example of the famous Bogoliubov $1/q^2$ theorem, which states that the Green function in long wave limit has $1/q^2$ divergence as $q\rightarrow0$.
 The above Josephson relation in superfluid system can be viewed as a manifestation of Bogoliubov's ``$1/q^2$" theorem  in systems  with spontaneously broken symmetries \cite{Bogoliubov}.
 It also has close connections with the absence of long ranged order (e.g., condensation) at finite temperature in one and two  dimensions \cite{Baym,Holzmann2003}.
 % The relation has been re-derived  with various methods \cite {Griffin,Holzmann,Muller}.
%with spontaneously broken symmetry

%Bose-Einstein condensation (BEC) and superfluidity are two closely related phenomenons in Bose quantum liquid at low temperature Generally speaking, condensate density $n_0$ is not equal to superfluid density $\rho_s$ in interacting bosons .
%B. D. Josephson derived a remarkable relation between the superfluid density $\rho_s$ and condensate density $n_0$ (the so-called Josephson relation)

Its possible generalization in two-component fermion
superfluid has been firstly investigated by Taylor \cite{Taylor}.
Using auxiliary-field approach, Dawson, et al. also derived  a Josephson relation which is suitable for both bosonic and fermion superfluid \cite{Dawson}, which is
\begin{align}
\rho_s=-{\rm lim_{q\rightarrow0}}\frac{4 \Delta^{2}m}{q^2 G_{II}(\textbf{q},0)},
\label{bas}
\end{align}
where superfluid order parameter (pairing gap in superconductor) $\Delta=\langle \psi_{\downarrow}(\textbf{r})\psi_{\uparrow}(\textbf{r})\rangle$ for usual two-component fermion superfluid. $G_{II}(\textbf{q},\omega)=\sum_n[\frac{|\langle0|\Delta_q|n\rangle|^2}{\omega-\omega_{n0}}-\frac{|\langle0|\Delta^{\dag}_q|n\rangle|^2}{\omega+\omega_{n0}}]$ is two-particle Green function for pairing states, $\Delta_{\textbf{q}}=\sum_{k}\psi_{\downarrow\textbf{q+k}}\psi_{\uparrow\textbf{-k}}$ is fluctuation operator of order parameter and $\psi_{\sigma\textbf{k}}$ is field operator in momentum space. $\omega_{n0}=E_n-E_0$, $E_n$ and $|n\rangle$ are system eigenenergy and eigenstates, respectively.
It is found that the superfluid density is determined by superfluid order parameter and the behaviors of two-particle Green function at long wave length limit. In comparison with bosonic superfluid, the above formula shows that the two-particle Green function replace the corresponding single-particle Green function of bosonic superfluid. In addition,
 pairing gap $\Delta$ plays the roles of order parameter in fermion superfluid.

The superfluid properties in multi-component (or multi-band) fermion system have been attracted a great interests \cite{ Modawi1997,Wu2003,Peotta2015,Yerin2019,Iskin2020}.
The exotic pairing mechanism in Fermi gas with SU(N) invariant interaction has been proposed \cite{Honerkamp2004,He2006,Zhai2007,Rapp2007,Catelani2008,Ozawa2010,Taie2010,Yip2011}, dependent on interaction and chemical potential, which can show coexistence of superfluid and magnetism \cite{Cherng2007}.
Another interesting  example of multi-band fermion system is twisted bilayer graphene \cite{Bistritzer2011,Morell2010}. It is shown that, there exists superconductivity  \cite{Cao2018} in this system. Its superfluid weight (which is superfluid density up to a constant) have been investigated intensively \cite{Hazra2019,Julku2020,Julku2016,Xie2020,Hu2019}.
Inspired by above studies, in this paper we investigate the superfluid density (or superfluid weight) for multi-component fermion superfluid. A generalized formula of Josephson relation for multi-component bosons has been given in Ref.\cite{zhangyicai2018}. A natural question arises: does there exist a similar relation for a multi-component fermion superfluid or superconductor?

 In this paper, we would give a generalized Josephson relation for a multi-component fermion superfluid or superconductor system. It is found that we can take a similar method as done in bosonic system to get the general results for fermions.
 In specific, the superfluid density can be expressed in terms of  two-particle Green functions.   Further more, when there is only one gapless collective mode, the superfluid density is determined by the superfluid order parameter and the trace of  two-particle Green function. In addition, in the presence of inverse symmetry, the two-particle Green function is directly related to fluctuation matrix of order parameters. Further more, it is found that the generalized Josephson relation can be also applied to multi-band lattice system.

The paper is organized as follows. In Section. \textbf{II}, we give a derivation of Josephson relation in usual two-component fermion superfluid.  The Josephson relation is generalized to a multi-component fermion superfluid in Sec. \textbf{III}.
  Two different formulas  for Josephson relations [Eqs.(\ref{spdensity2}) and (\ref{spdensity3})] are given in Sec. \textbf{IV}. In addition, we take Haldane model as example to illustrate above results in Sec. \textbf{V}.
   A summary is given in Sec. \textbf{VI}.

\section{Josephson relation in usual two component fermion superfluid}
In this part, similarly as bosonic case \cite{zhangyicai2018}, we give  another derivation, which is different from that of Taylor \cite{Taylor} and Dawson et.al \cite{Dawson}.
In the rest of paper, we take $m=\hbar=V=1$, where $m$ is particle mass, $V$ is volume of system.
%The Josephson relation gives a connection of superfluid density $\rho_s$, superfluid order parameter $\Delta_0$ and two-particle Green function at zero frequency and low momentum limit \cite{Taylor,Dawson}
%\begin{align}
%\rho_s=-{\rm lim_{q\rightarrow0}}\frac{4\Delta^{2}_{0}}{q^2 G_{II}(\textbf{q},0)},
%\end{align}
%where two-particle retarded Green's function is
%\begin{align}
%G_{II}(\textbf{q},\omega+i\eta)=\sum_n[\frac{|\langle0|\Delta_\textbf{q}|n\rangle|^2}{\omega+i\eta-\omega_{n0}}-\frac{|\langle0|\Delta^{\dag}_\textbf{q}|n\rangle|^2}{\omega+i\eta+\omega_{n0}}],
%\end{align}
%with order parameter fluctuation operator of momentum space $\Delta_\textbf{q}=\sum_{k}\psi_{\downarrow\textbf{q+k}}\psi_{\uparrow\textbf{-k}}$, $\Delta^{\dag}_\textbf{q}$ and $\psi_{\sigma\textbf{k}}$ is field operator in momentum space. $\omega_{n0}=E_n-E_0$, $E_n$ and $|n\rangle$ are system eigenenergy and eigenstates, respectively.
%
Here we outline how to get the above relation Eq.(\ref{bas}) in usual two-component fermion system. Firstly we know that, when the field operators undergo a phase variation \cite{Lifshitz}, namely,
 \begin{align}
 \psi_{\sigma}(\textbf{r})\rightarrow e^{i\delta\theta(\textbf{r})}\psi_{\sigma}(\textbf{r}),
 \label{field}
\end{align}
where $\delta\theta$ is a real function which denotes the phase variation, the superfluid order parameter would also has a phase variation.
The superfluid order parameter (or pairing gap in superconductor) for two-component fermions is
 \begin{align}
 \Delta(r)\equiv g\langle\psi_{\downarrow}(\textbf{r})\psi_{\uparrow}(\textbf{r})\rangle=\Delta.
 \label{orderparameter}
\end{align}
 where $g$ is interaction strength between two different (spin) components. In the following, we would drop the interaction parameter $g$ for simplification.
Substituting Eq.(\ref{field}) into Eq.(\ref{orderparameter}), we get the variation of $\Delta(\textbf{r})$
\begin{align}
 &\delta \Delta(\textbf{r})= \langle\psi_{\downarrow}(\textbf{r})\psi_{\uparrow}(\textbf{r})\rangle e^{2i\delta\theta(\textbf{r})}-\langle\psi_{\downarrow}(\textbf{r})\psi_{\uparrow}(\textbf{r})\rangle,\notag\\
 &\simeq 2i\langle\psi_{\downarrow}(\textbf{r})\psi_{\uparrow}(\textbf{r})\rangle \delta\theta(\textbf{r})=2i\Delta \delta\theta(\textbf{r}).
\label{deltaphi}
\end{align}

On the other hand,  the current (superflow) density is also related to the phase, namely
 \begin{align}
 \delta\textbf{j}(\textbf{r})\equiv\rho_{s}\vec{\nabla}\delta\theta(\textbf{r})=\rho_{s}\delta\textbf{v}_s,
 \label{deltaj}
\end{align}
 where we define superfluid velocity  as gradient of phase variation, i.e., $\delta\textbf{v}_s\equiv\vec{\nabla} \delta\theta(\textbf{r})$ and superfluid density $\rho_s$ as the coefficient before $\delta\textbf{v}_s$ in the current $\delta\textbf{j}(\textbf{r})$. We will see the connection between the above two equations (eqs.(\ref{deltaj}) and (\ref{deltaphi})) would result in the Josephson relation.

In order to get the relationship between superfluid density $\rho_s$ and order parameter $\Delta$, similarly as bosonic case \cite{Baym,Ueda},  here we need add a perturbation which couples to  operator $\psi_{\downarrow}\psi_{\uparrow}$ and its adjoint $\psi^{\dag}_{\uparrow}\psi^{\dag}_{\downarrow}$, i.e.,
\begin{align}
& H'=\notag\\
&\!\!\int \!\! d^3\textbf{r}[e^{i(\textbf{q}\cdot \textbf{\textbf{r}}-\omega t)+\epsilon t}\xi \psi^{\dag}_{\uparrow}(\textbf{r})\psi^{\dag}_{\downarrow}(\textbf{r})\!\!+\!\!e^{-i(\textbf{q}\cdot \textbf{r}-\omega t)+\epsilon t}\xi^* \psi_{\downarrow}(\textbf{r})\psi_{\uparrow}(\textbf{r})],\notag\\
 &=\xi\Delta^{\dag}_{\textbf{q}}e^{-i\omega t+\epsilon t}+\xi^*\Delta_{\textbf{q}}e^{i\omega t+\epsilon t},
\end{align}
where $\xi$ is a small complex number, $\Delta_{\textbf{q}}=\sum_{k}\psi_{\downarrow\textbf{q+k}}\psi_{\uparrow\textbf{-k}}$ is fluctuation operator of order parameter and we use relation $\psi_\sigma(\textbf{r})=\sum_\textbf{k} \psi_{\sigma\textbf{k}}e^{i\textbf{k}\cdot \textbf{r}}$. Here we add an infinitesimal positive number $\epsilon\rightarrow0_+$ in the above exponential which corresponds choosing boundary condition that the perturbation is very slowly added to the system \cite{Pines1}.

We assume initially the system is in the ground state $|0\rangle$, and then slowly turn on perturbation $H'$, the wave function can be written as
\begin{align}
 &&\psi(t)=\sum_n a_n(t)e^{-iE_nt}\psi_n,
\end{align}
where $a_n(t\rightarrow -\infty)=\delta_{n,0}$ and $H\psi_n=E_n\psi_n$, $H$ is unperturbated Hamiltonian.
Using perturbation theory, we get
\begin{align}
 &&&\psi(t)\simeq \psi_0e^{-iE_0 t}+\sum_{n\neq0} a_n(t)e^{-iE_nt}\psi_n,\notag\\
 &&& a_n(t)=\frac{1}{i}\int_{-\infty}^{t} d\tau H'_{n0}(\tau)e^{i\omega_{n0}\tau}\notag\\
 &&&\!\!=\!\![\frac{\xi\langle n|\Delta^{\dag}_{\textbf{q}}|0\rangle e^{-i(\omega+i\eta-\omega_{n0})t}}{\omega+i\epsilon-\omega_{n0}}-\frac{\xi^*\langle n|\Delta_{\textbf{q}}|0\rangle e^{i(\omega-i\eta+\omega_{n0})t}}{\omega-i\epsilon+\omega_{n0}}],\notag\\
\end{align}
where $\omega_{n0}=E_n-E_0$.
The changes of order parameter $\langle\psi_{\downarrow}\psi_{\uparrow}(\textbf{r})\rangle$ and current $\textbf{j}(\textbf{r})$ are respectively
\begin{align}
 &&&\delta\Delta(\textbf{r})=\delta\langle\psi_{\downarrow}\psi_{\uparrow}(\textbf{r})\rangle\notag\\
 &&&=\!\!\xi e^{-i(\omega+i\epsilon)t}[\frac{\langle0|\psi_{\downarrow}\psi_{\uparrow}(\textbf{r})|n\rangle\langle n|\Delta^{\dag}_{\textbf{q}}|0\rangle}{\omega+i\epsilon-\omega_{n0}}\!\!-\!\!\frac{\langle0|\Delta^{\dag}_{\textbf{q}}|n\rangle\langle n|\psi_{\downarrow}\psi_{\uparrow}(\textbf{r})|0\rangle}{\omega+i\epsilon+\omega_{n0}}]\notag\\
 &&&\!\!+\!\!\xi^* e^{i(\omega-i\epsilon)t}[\frac{\langle0|\Delta_{\textbf{q}}|n\rangle\langle n|\psi_{\downarrow}\psi_{\uparrow}(\textbf{r})|0\rangle}{\omega-i\epsilon-\omega_{n0}}\!\!-\!\!\frac{\langle0|\psi_{\downarrow}\psi_{\uparrow}(\textbf{r})|n\rangle\langle n|\Delta_{\textbf{q}}|0\rangle}{\omega-i\epsilon+\omega_{n0}}],\notag\\
 &&&\delta \textbf{j}(\textbf{r})=\xi e^{-i(\omega+i\eta)t}[\frac{\langle0|\textbf{j}(\textbf{r})|n\rangle\langle n|\Delta^{\dag}_{\textbf{q}}|0\rangle}{\omega+i\epsilon-\omega_{n0}}-\frac{\langle0|\Delta^{\dag}_{\textbf{q}}|n\rangle\langle n|\textbf{j}(\textbf{r})|0\rangle}{\omega+i\epsilon+\omega_{n0}}]\notag\\
 &&&+\xi^* e^{i(\omega-i\epsilon)t}[\frac{\langle0|\Delta_{\textbf{q}}|n\rangle\langle n|\textbf{j}(\textbf{r})|0\rangle}{\omega-i\epsilon-\omega_{n0}}-\frac{\langle0|\textbf{j}(\textbf{r})|n\rangle\langle n|\Delta_{\textbf{q}}|0\rangle}{\omega-i\epsilon+\omega_{n0}}].\notag\\
\end{align}
In the following, we assume the system has translational invariance and momentum is a good quantum number. So every eigenstate $|n\rangle$ has a definite momentum, e.g., $\textbf{q}_n$ and $\textbf{P}|n\rangle=\textbf{q}_n|n\rangle$ with momentum operator $\textbf{P}=\sum_{\sigma \textbf{k}} \textbf{k}\psi_{\sigma\textbf{k}}^{\dag}\psi_{\sigma\textbf{k}}$.
On other hand, from commutation relations
\begin{align}
 &&[\textbf{P},\Delta_{\textbf{q}}^{\dag}]|n\rangle=\{\textbf{P}\Delta_{\textbf{q}}^{\dag}-\Delta_{\textbf{q}}^{\dag}\textbf{P}\}|n\rangle=\textbf{q}\Delta_{\textbf{q}}^{\dag}|n\rangle,\notag\\
 &&[\textbf{P},\Delta_{\textbf{q}}]|n\rangle=\{\textbf{P}\Delta_{\textbf{q}}-\Delta_{\textbf{q}}\textbf{P}\}|n\rangle=-\textbf{q}\Delta_{\textbf{q}}|n\rangle,
\end{align}
we see $\Delta_{\textbf{q}}^{\dag}|n\rangle$ and $\Delta_{\textbf{q}}|n\rangle$ also have definite momenta which are  $\textbf{q}_n+\textbf{q}$ and $\textbf{q}_n-\textbf{q}$, respectively.
Using $\psi_{\downarrow}(\textbf{r})\psi_{\uparrow}(\textbf{r})=\sum_\textbf{q}\Delta_\textbf{q} e^{i\textbf{q}\cdot\textbf{r}}$ and $\langle 0|\Delta_{\textbf{q}'}|n\rangle\langle n|\Delta_{\textbf{q}}^{\dag}|0\rangle=\delta_{\textbf{q},\textbf{q}'}|\langle 0|\Delta_\textbf{q}|n\rangle|^2$, $\langle 0|\Delta_{\textbf{q}}^{\dag}|n\rangle\langle n|\Delta_{\textbf{q}'}|0\rangle=\delta_{\textbf{q},\textbf{q}'}|\langle 0|\Delta_{\textbf{q}}^{\dag}|n\rangle|^2$, $\langle 0|\Delta_{\textbf{q}}|n\rangle\langle n|\Delta_{\textbf{q}'}|0\rangle=\delta_{\textbf{q},-\textbf{q}'}\langle 0|\Delta_{\textbf{q}}|n\rangle\langle n|\Delta_{-\textbf{q}}|0\rangle$, $\langle 0|\Delta_{\textbf{q}'}|n\rangle\langle n|\Delta_{\textbf{q}}|0\rangle=\delta_{\textbf{q},-\textbf{q}'}\langle 0|\Delta_{-\textbf{q}}|n\rangle\langle n|\Delta_{\textbf{q}}|0\rangle$,
the variation of order parameter can be written as
\begin{align}
&&\delta\langle\psi_{\downarrow}\psi_{\uparrow}(\textbf{r})\rangle=\xi e^{i\textbf{q}\cdot \textbf{r}-i(\omega+i\epsilon)t} G_{II}(\textbf{q},\omega+i\epsilon)\notag\\
&& +\xi^* e^{-i\textbf{q}\cdot \textbf{r}+i(\omega-i\epsilon)t}F_{II}(\textbf{q},\omega-i\epsilon),
\end{align}
 where
 \begin{align}
&&&G_{II}(\textbf{q},\omega+i\epsilon)=\sum_n[\frac{|\langle0|\Delta_\textbf{q}|n\rangle|^2}{\omega+i\epsilon-\omega_{n0}}
 -\frac{|\langle0|\Delta^{\dag}_\textbf{q}|n\rangle|^2}{\omega+i\epsilon+\omega_{n0}}],\notag\\
 &&&F_{II}(\textbf{q},\omega-i\epsilon)\notag\\
 &&&=\sum_n[\frac{\langle0|\Delta_{\textbf{q}}|n\rangle\langle n|\Delta_{-\textbf{q}}|0\rangle}{\omega-i\epsilon-\omega_{n0}}-\frac{\langle0|\Delta_{-\textbf{q}}|n\rangle\langle n|\Delta_{\textbf{q}}|0\rangle}{\omega-i\epsilon+\omega_{n0}}],
 \end{align}
 is two-particle normal (anomalous) Green function for pairing states \cite{Nozieres,Abrikosov}.

Taking zero-frequency of $\omega\pm i \epsilon =0$ limit,
\begin{align}
 &&\delta\langle\psi_{\downarrow}\psi_{\uparrow}(\textbf{r})\rangle=\xi e^{i\textbf{q}\cdot \textbf{r}} G_{II}(\textbf{q},0)+\xi^* e^{-i\textbf{q}\cdot \textbf{r}}F_{II}(\textbf{q},0).
\end{align}
For two-component neutral fermions, the order parameter $\Delta(\textbf{r})=\langle \psi_{\downarrow}(\textbf{r})\psi_{\uparrow}(\textbf{r})\rangle=\Delta$ can be taken as a real number,  and the low energy collective excitation is Anderson-Bogoliubov phonon. Similarly as bosonic case \cite{zhangyicai2018}, it can be shown that $F_{II}(\textbf{q},0)=-G_{II}(\textbf{q},0)$ as $q\rightarrow0$ (see Sec.\textbf{IV}), so finally
\begin{align}
&&\delta\Delta(\textbf{r})=G_{II}(\textbf{q},0)[\xi e^{i\textbf{q}\cdot \textbf{r}} -\xi^* e^{-i\textbf{q}\cdot \textbf{r}}],\notag\\
&&=2i\alpha G_{II}(\textbf{q},0)sin(\textbf{q}\cdot \textbf{r}+\phi),
\label{eq15}
\end{align}
where we take $\xi\equiv\alpha e^{i\phi}$ with amplitude $\alpha$ and phase $\phi$.

Similarly using commutation relation
\begin{align}
 &&[\textbf{P}_i,\textbf{j}_{\textbf{q}j}]=-\textbf{q}_i \textbf{j}_{\textbf{q}j},
\end{align}
where indices $i,j=x,y,z$, current fluctuation operator $\textbf{j}_{\textbf{q}}=\sum_{\sigma\textbf{k}} [\textbf{k}+\textbf{q}/2]\psi^{\dag}_{\sigma\textbf{k}}\psi_{\sigma\textbf{k}+\textbf{q}}$, and the translational invariance, we conclude $\textbf{j}_{\textbf{q}}|n\rangle$ also has a definite momentum $\textbf{q}_n-\textbf{q}$. Using the fact of $\textbf{j}(\textbf{r})=\sum_\textbf{q} \textbf{j}_\textbf{q} e^{i\textbf{q}\cdot\textbf{r}}$ and $\langle 0|\textbf{j}_{\textbf{q}'}|n\rangle\langle n|\Delta_{\textbf{q}}^{\dag}|0\rangle=\delta_{\textbf{q},\textbf{q}'}\langle 0|\textbf{j}_\textbf{q}|n\rangle\langle n|\Delta_{\textbf{q}}^{\dag}|0\rangle$, $\langle 0|\Delta_{\textbf{q}}^{\dag}|n\rangle\langle n|\textbf{j}_{\textbf{q}'}|0\rangle=\delta_{\textbf{q},\textbf{q}'}\langle 0|\Delta_{\textbf{q}}^{\dag}|n\rangle\langle n|\textbf{j}_{\textbf{q}}|0\rangle$, the variation of current is
\begin{align}
\delta \textbf{j}(\textbf{r})=\xi e^{i\textbf{q}\cdot \textbf{r}-i(\omega+i\epsilon)t}\textbf{B}(\textbf{q},\omega+i\epsilon)+h.c.,
\end{align}
where $h.c.$ denotes Hermitian (complex) conjugate and
 \begin{align}
\textbf{B}(\textbf{q},\omega+i\epsilon)\equiv\sum_n[\frac{\langle0|\textbf{j}_q|n\rangle\langle n|\Delta^{\dag}_{\textbf{q}}|0\rangle}{\omega+i\epsilon-\omega_{n0}}-\frac{\langle0|\Delta^{\dag}_\textbf{q}|n\rangle\langle n|\textbf{j}_\textbf{q}|0\rangle}{\omega+i\epsilon+\omega_{n0}}].
\end{align}

When $\omega\pm i\epsilon=0$,
\begin{align}
\delta\textbf{j}(\textbf{r})=[\xi e^{i\textbf{q}\cdot \textbf{r}}\textbf{B}(\textbf{q},0) +h.c.].
\end{align}

Using continuity equation $\frac{\partial\rho(\textbf{r},t)}{\partial t}+\vec{\nabla}\cdot\textbf{ j}(\textbf{r},t)=0$, $\omega_{n0}(\rho_\textbf{q})_{0n}=\textbf{q}\cdot(\textbf{j}_{\textbf{q}})_{0n}$ and $\omega_{n0}(\rho_\textbf{q})_{n0}=-\textbf{q}\cdot(\textbf{j}_{\textbf{q}})_{n0}$, we can obtain
\begin{align}
&\textbf{q}\cdot\textbf{B}(\textbf{q},0)=-\sum_n[\langle0|\rho_\textbf{q}|n\rangle\langle n|\Delta^{\dag}_{\textbf{q}}|0\rangle-\langle0|\Delta^{\dag}_\textbf{q}|n\rangle\langle n|\rho_\textbf{q}|0\rangle],\notag\\
&=-\langle0|[\rho_\textbf{q},\Delta^{\dag}_\textbf{q}]|0\rangle=-2\langle0|\Delta^{\dag}_{\textbf{q}=\textbf{0}}|0\rangle=-2\Delta,
\end{align}
where density fluctuation operator $\rho_\textbf{q}=\sum_{\sigma\textbf{k}} \psi^{\dag}_{\sigma \textbf{k}}\psi_{\sigma\textbf{k}+\textbf{q}}$ and we use the fact that $\langle\Delta^{\dag}_{\textbf{q}=0}\rangle=\Delta^*=\Delta$. So
\begin{align}
&&\textbf{q}\cdot \delta \textbf{j}(r)=-2[\xi e^{i\textbf{q}\cdot \textbf{r}}\Delta+h.c.].
\label{yiban}
\end{align}
For isotropic system, further assuming $\textbf{q}\parallel \textbf{B}\propto \delta \textbf{j}$ and using Eq.(\ref{eq15}), so we get
\begin{align}
&\delta \textbf{j}(\textbf{r})=-\frac{2\textbf{q}\Delta}{q^2}[\xi e^{i\textbf{q}\cdot \textbf{r}}+\xi^* e^{-i\textbf{q}\cdot \textbf{r}}],\notag\\
&=-2\Delta \frac{\textbf{q}}{q^2} 2\alpha cos(\textbf{q}\cdot \textbf{r}+\phi)=-\frac{2\Delta}{q^2} \frac{\vec{\nabla }\delta \Delta(\textbf{r})}{iG_{II}(\textbf{q},0)}.
\end{align}
Here we further use Eq.(\ref{deltaphi}), and then get
\begin{align}
&&\delta \textbf{j}(\textbf{r})= -\frac{4\Delta^{2}}{q^2} \frac{\vec{\nabla }\delta \theta(\textbf{r})}{G_{II}(\textbf{q},0)}=-\frac{4\Delta^{2}}{q^2} \frac{ \delta \textbf{v}_s}{G_{II}(\textbf{q},0)}.
\end{align}
Using Eq.(\ref{deltaj}), i.e., $\delta \textbf{j}(\textbf{r})\equiv\rho_s  \delta \textbf{v}_s$, the Josephson relation for usual two-component Fermions is obtained
\begin{align}
\rho_s=-{\rm lim_{q\rightarrow0}}\frac{4\Delta^{2}}{q^2 G_{II}(\textbf{q},0)} ,\label{Josephson}
\end{align}
which is consistent with the Taylor's  \cite{Taylor} and Dawson et. al.'s  \cite{Dawson} results.

\section{General Josephson relation for multi-component fermions }
For a multi-component (or multi-band) fermions, the superfluid order parameter $\Delta_{\alpha\beta}$ can be written as
\begin{align}
 \Delta_{\alpha\beta}=\langle\psi_{\alpha}(\textbf{r})\psi_{\beta}(\textbf{r})\rangle,
\label{H1}
\end{align}
where $\psi_{\alpha(\beta)}$ is the field operator for $\alpha(\beta)$-th component.
 The above equation indicates that the number of superfluid order parameter can be an arbitrary integer $m\geq1$ ($m\in Z$) in a multi-component fermion superfluid. In such a case, we need a general perturbation Hamiltonian
\begin{align}
 &H'=\int d^3r\{e^{i(\textbf{q}\cdot \textbf{r}-\omega t+\epsilon t)} \Delta^{\dag}(\textbf{r}).\xi+e^{-i(\textbf{q}\cdot\textbf{ r}-\omega t+\epsilon t)}\xi^{\dag} .\Delta(\textbf{r})\},\notag\\
 &=\Delta^{\dag}_{\textbf{q}}.\xi e^{-i\omega t+\epsilon t}+\xi^{\dag}.\Delta_{\textbf{q}}e^{i\omega t+\epsilon t},
\label{H1}
\end{align}
where we relabel the order parameter with $\Delta_{i}(i=1,2,...,m)$ and introduce column vectors $\Delta(\textbf{r})=\{\Delta_1(\textbf{r}),\Delta_2(\textbf{r}),...,\Delta_m(\textbf{r}) \}^\textbf{t}$, $\xi=\{ \xi_1,\xi_2,...,\xi_m \}^\textbf{t}$ and $\{...\}^t$ denotes matrix transpose.
Similarly, using perturbation theory, we can get the variation of order parameter
\begin{align}
 &&&\delta\Delta_{\sigma}(\textbf{r})=\sum_{\sigma'}[e^{i\textbf{q}\cdot \textbf{r}-i(\omega+i\epsilon)t}  G_{II\sigma\sigma'}(\textbf{q},\omega+i\epsilon)\xi_{\sigma'}\notag\\
 &&&+e^{-i\textbf{q}\cdot \textbf{r}+i(\omega-i\epsilon)t} F_{II\sigma'\sigma}(\textbf{q},\omega-i\epsilon) \xi_{\sigma'}^{*}],
 \label{definition}
\end{align}
where
\begin{align}
 &&&G_{II\sigma\sigma'}(\textbf{q},\omega+i\epsilon)\notag\\
 &&&=\sum_n[\frac{\langle0|\Delta_{\sigma \textbf{q}}|n\rangle\langle n|\Delta^{\dag}_{\sigma' \textbf{q}}|0\rangle}{\omega+i\epsilon-\omega_{n0}}-\frac{\langle0|\Delta^{\dag}_{\sigma' \textbf{q}}|n\rangle\langle n|\Delta_{\sigma \textbf{q}}|0\rangle}{\omega+i\epsilon+\omega_{n0}}],\notag\\
 &&&F_{II\sigma'\sigma}(\textbf{q},\omega-i\epsilon)\notag\\
 &&&=\sum_n[\frac{\langle0|\Delta_{\sigma',\textbf{q}}|n\rangle\langle n|\Delta_{\sigma,-\textbf{q}}|0\rangle}{\omega-i\epsilon-\omega_{n0}}-\frac{\langle0|\Delta_{\sigma,-\textbf{q}}|n\rangle\langle n|\Delta_{\sigma',\textbf{q}}|0\rangle}{\omega-i\epsilon+\omega_{n0}}],
 \label{definition}
\end{align}
are two-particle normal (anomalous) Green function matrix elements and indices $\sigma (\sigma')=1,2,...,m$.

Similarly as above section, one gets
\begin{align}
\delta \textbf{j}(\textbf{r})=e^{i\textbf{q}\cdot\textbf{r}-i(\omega+i\epsilon)t} \textbf{B}(q,\omega+i\epsilon).\xi+h.c.,
\end{align}
and
\begin{align}
 &&&\textbf{B}_{\sigma'}(q,\omega+i\epsilon)\notag\\
 &&&=\sum_n[\frac{\langle0|\textbf{j}_\textbf{q}|n\rangle\langle n|\Delta_{\sigma'\textbf{q}}^{\dag}|0\rangle}{\omega+i\epsilon-\omega_{n0}}-\frac{\langle0|\Delta_{\sigma'\textbf{q}}^{\dag}|n\rangle\langle n|\textbf{j}_\textbf{q}|0\rangle}{\omega+i\epsilon+\omega_{n0}}].\notag\\
\end{align}

When $\omega\pm i\epsilon= 0$,
\begin{align}
 &&&\delta\langle\Delta(\textbf{r})\rangle=e^{i\textbf{q}\cdot\textbf{r}} G_{II}(\textbf{q},0).\xi+e^{-i\textbf{q}\cdot\textbf{r}}  F_{II}^{\textbf{t}}(\textbf{q},0).\xi^{*},\notag\\
 &&&\delta \textbf{j}(\textbf{r})=e^{i\textbf{q}\cdot\textbf{r}} \textbf{B}(\textbf{q},0).\xi+h.c.
 \label{current}
\end{align}
Similarly, using continuity equations and assuming $\textbf{q}\parallel \textbf{B}\propto \delta \textbf{j}$ for isotropic system, we get
\begin{align}
 \textbf{B}(\textbf{q},0)=-\frac{2\textbf{q}}{q^2}\{\Delta^{*}_{1},\Delta^{*}_{2},...,\Delta^{*}_{m}\}.
\end{align}

Introducing $x(\textbf{r})=e^{i\textbf{q}\cdot\textbf{r}}\{\xi_{1},\xi_{2},...,\xi_{m} \}^\textbf{t}$, $\delta \Delta(\textbf{r})=\{\delta\Delta_{1}(\textbf{r}), \delta\Delta_{2}(\textbf{r}),...,\delta\Delta_{m}(\textbf{r})\}^\textbf{t}$,  the above equations can be written as
\begin{align}
&&&\left( \!\!\!                %×óÀšºÅ
 \begin{array}{cccc}   %žÃŸØÕóÒ»¹²3ÁÐ£¬Ã¿Ò»ÁÐ¶ŒŸÓÖÐ·ÅÖÃ
   \delta \langle\Delta(\textbf{r})\rangle    \\  %µÚÒ»ÐÐÔªËØ
   \delta \langle\Delta(\textbf{r})\rangle^{*} \\  %µÚ¶þÐÐÔªËØ
\end{array}\!\!\!\right)\!\!=\!\!\left( \!\!\!                %×óÀšºÅ
 \begin{array}{cccc}   %žÃŸØÕóÒ»¹²3ÁÐ£¬Ã¿Ò»ÁÐ¶ŒŸÓÖÐ·ÅÖÃ
   G_{II}(\textbf{q},0)  & F_{II}^{\textbf{t}}(\textbf{q},0) \\  %µÚÒ»ÐÐÔªËØ
   F_{II}^{\textbf{t}*}(\textbf{q},0)& G_{II}^{*}(\textbf{q},0)\\  %µÚ¶þÐÐÔªËØ
\end{array}\!\!\!\right)_{2m\times2m}.\left( \!\!\!                %×óÀšºÅ
 \begin{array}{cccc}   %žÃŸØÕóÒ»¹²3ÁÐ£¬Ã¿Ò»ÁÐ¶ŒŸÓÖÐ·ÅÖÃ
   x   \\  %µÚÒ»ÐÐÔªËØ
   x^* \\  %µÚ¶þÐÐÔªËØ
\end{array}\!\!\!\right)\notag\\
&&&\equiv\textbf{G}_{\textbf{II}}.\left( \!\!\!                %×óÀšºÅ
 \begin{array}{cccc}   %žÃŸØÕóÒ»¹²3ÁÐ£¬Ã¿Ò»ÁÐ¶ŒŸÓÖÐ·ÅÖÃ
   x   \\  %µÚÒ»ÐÐÔªËØ
   x^* \\  %µÚ¶þÐÐÔªËØ
\end{array}\!\!\!\right)
=2i\delta\theta\left( \!\!\!                %×óÀšºÅ
 \begin{array}{cccc}   %žÃŸØÕóÒ»¹²3ÁÐ£¬Ã¿Ò»ÁÐ¶ŒŸÓÖÐ·ÅÖÃ
   \Delta   \\  %µÚÒ»ÐÐÔªËØ
   -\Delta^* \\  %µÚ¶þÐÐÔªËØ
\end{array}\!\!\!\right)
\label{basic}
\end{align}

and
\begin{align}
&&\delta \textbf{j}(\textbf{r})\!\!=\!\!-\frac{2\textbf{q}}{q^2}\left(                %×óÀšºÅ
 \begin{array}{cccc}   %žÃŸØÕóÒ»¹²3ÁÐ£¬Ã¿Ò»ÁÐ¶ŒŸÓÖÐ·ÅÖÃ
   \Delta^{t*}  &\Delta^{t} \\  %µÚÒ»ÐÐÔªËØ
\end{array}\right).\!\!\left( \!\!\!                %×óÀšºÅ
 \begin{array}{cccc}   %žÃŸØÕóÒ»¹²3ÁÐ£¬Ã¿Ò»ÁÐ¶ŒŸÓÖÐ·ÅÖÃ
   (I)_{m\times m}  & 0_{m\times m}  \\  %µÚÒ»ÐÐÔªËØ
   0_{m\times m} &  (I)_{m\times m}\\  %µÚ¶þÐÐÔªËØ
\end{array}\!\!\!\right).\left(                 %×óÀšºÅ
 \begin{array}{cccc}   %žÃŸØÕóÒ»¹²3ÁÐ£¬Ã¿Ò»ÁÐ¶ŒŸÓÖÐ·ÅÖÃ
   x   \\  %µÚÒ»ÐÐÔªËØ
   x^* \\  %µÚ¶þÐÐÔªËØ
\end{array}\!\!\!\right),
\label{basic2}
\end{align}
where $(I)_{m\times m}$ is a $m\times m$ identity matrix and we define coefficient matrix
 \begin{align}
 \textbf{G}_{\textbf{II}}\equiv\!\!\left( \!\!\!                %×óÀšºÅ
 \begin{array}{cccc}   %žÃŸØÕóÒ»¹²3ÁÐ£¬Ã¿Ò»ÁÐ¶ŒŸÓÖÐ·ÅÖÃ
   G_{II}(\textbf{q},0)  & F_{II}^\textbf{t}(\textbf{q},0) \\  %µÚÒ»ÐÐÔªËØ
   F_{II}^{t*}(\textbf{q},0)& G_{II}^{*}(\textbf{q},0)\\  %µÚ¶þÐÐÔªËØ
\end{array}\!\!\!\right)_{2m\times 2m} ,
\end{align}
 which is a $2m\times2m$ matrix and one should not confuse with normal Green function $G_{II}(\textbf{q},0)$, which a $m\times m$ matrix.
 %The two eqs. (\ref{basic}) and (\ref{basic2}) are basic equations in this work. In the following, most of discussions are based on them.
If $\textbf{G}_{\textbf{II}}$ has inverse (determinant $Det|\textbf{G}_{\textbf{II}}|\neq0$),
 using $\textbf{q} x(\textbf{r})=-i \vec{\nabla } x$, $\textbf{q} x^*(\textbf{r})=i \vec{\nabla } x^*$ and eqs. (\ref{basic}) and (\ref{basic2}),
we get
\begin{eqnarray}
&& \delta \textbf{j}(\textbf{r})=-\frac{4\vec{\nabla}\delta \theta(\textbf{r})}{q^2}(\Delta^{t*},\Delta^{t})\notag\\
&&.\!\!\left( \!\!\!                %×óÀšºÅ
 \begin{array}{cccc}   %žÃŸØÕóÒ»¹²3ÁÐ£¬Ã¿Ò»ÁÐ¶ŒŸÓÖÐ·ÅÖÃ
  I  & 0 \\  %µÚÒ»ÐÐÔªËØ
   0&  -I\\  %µÚ¶þÐÐÔªËØ
\end{array}\!\!\!\right).\textbf{G}_{\textbf{II}}^{-1}.\!\!\left( \!\!\!                %×óÀšºÅ
 \begin{array}{cccc}   %žÃŸØÕóÒ»¹²3ÁÐ£¬Ã¿Ò»ÁÐ¶ŒŸÓÖÐ·ÅÖÃ
  I  & 0 \\  %µÚÒ»ÐÐÔªËØ
   0&  -I\\  %µÚ¶þÐÐÔªËØ
\end{array}\!\!\!\right).\left(
 \begin{array}{cccc}   %žÃŸØÕóÒ»¹²3ÁÐ£¬Ã¿Ò»ÁÐ¶ŒŸÓÖÐ·ÅÖÃ
   \Delta \\  %µÚÒ»ÐÐÔªËØ
   \Delta^*\\  %µÚ¶þÐÐÔªËØ
\end{array}\right),\notag\\
&&=-\frac{4\delta \textbf{v}_s}{q^2}(\Delta^{t*},\Delta^t).\notag\\
&&.\!\!\left( \!\!\!                %×óÀšºÅ
 \begin{array}{cccc}   %žÃŸØÕóÒ»¹²3ÁÐ£¬Ã¿Ò»ÁÐ¶ŒŸÓÖÐ·ÅÖÃ
 I  & 0 \\  %µÚÒ»ÐÐÔªËØ
   0&  -I\\  %µÚ¶þÐÐÔªËØ
\end{array}\!\!\!\right).\textbf{G}_{\textbf{II}}^{-1}.\!\!\left( \!\!\!                %×óÀšºÅ
 \begin{array}{cccc}   %žÃŸØÕóÒ»¹²3ÁÐ£¬Ã¿Ò»ÁÐ¶ŒŸÓÖÐ·ÅÖÃ
  I  & 0 \\  %µÚÒ»ÐÐÔªËØ
   0&  -I\\  %µÚ¶þÐÐÔªËØ
\end{array}\!\!\!\right).\left(
 \begin{array}{cccc}   %žÃŸØÕóÒ»¹²3ÁÐ£¬Ã¿Ò»ÁÐ¶ŒŸÓÖÐ·ÅÖÃ
   \Delta \\  %µÚÒ»ÐÐÔªËØ
   \Delta^*\\  %µÚ¶þÐÐÔªËØ
\end{array}\right).
\end{eqnarray}

Similarly  using Eq. (\ref{deltaj}), i.e., $\delta \textbf{j}(\textbf{r})\equiv\rho_s \delta \textbf{v}_s$, we get a general Josephson relation for fermion superfluid
\begin{eqnarray}
&& \rho_s={\rm lim_{q\rightarrow0}}\frac{-4}{q^2}(\Delta^{t*},\Delta^t).\!\!\left( \!\!\!                %×óÀšºÅ
 \begin{array}{cccc}   %žÃŸØÕóÒ»¹²3ÁÐ£¬Ã¿Ò»ÁÐ¶ŒŸÓÖÐ·ÅÖÃ
  I  & 0 \\  %µÚÒ»ÐÐÔªËØ
   0&  -I\\  %µÚ¶þÐÐÔªËØ
\end{array}\!\!\!\right).\textbf{G}_{\textbf{II}}^{-1}.\!\!\left( \!\!\!                %×óÀšºÅ
 \begin{array}{cccc}   %žÃŸØÕóÒ»¹²3ÁÐ£¬Ã¿Ò»ÁÐ¶ŒŸÓÖÐ·ÅÖÃ
  I  & 0 \\  %µÚÒ»ÐÐÔªËØ
   0&  -I\\  %µÚ¶þÐÐÔªËØ
\end{array}\!\!\!\right)\!\!.\!\!\left(\!\!
 \begin{array}{cccc}   %žÃŸØÕóÒ»¹²3ÁÐ£¬Ã¿Ò»ÁÐ¶ŒŸÓÖÐ·ÅÖÃ
   \Delta \\  %µÚÒ»ÐÐÔªËØ
   \Delta^*\\  %µÚ¶þÐÐÔªËØ
\end{array}\!\!\right).\notag\\
\label{spdensity}
\end{eqnarray}
 The Eq. (\ref{spdensity}) gives a way to calculate the superfluid density in terms of two-particle Green functions, which is also the main result in this work.
 % From the above equation, we see the generalized Josephson relation also gives a connection between superfluid density $\rho_s$, order parameter $\Delta$ and Green function.

We should remark that  even though Eq.(\ref{spdensity}) is obtained in a translational invariant system (the momentum is a good quantum number), the above formula for superfluid density can be applied equally to lattice system as long as the superfluid states has lattice translation symmetry (see also Sec.\textbf{V}). This is because, we know  for lattice system, the engenstates can be classified by lattice momentum rather than true momentum. In the above derivation, on the one hand, we need to replace the momentum by the lattice momentum. However, on the other hand, the superfluid density may show some anisotropy in lattice system. We can use a similar method as done in bosonic system \cite{zhangyicai2018} to deal with the anisotropy in lattice. In such a case, the superfluid density is usually a second order tensor and depends on direction of $\textbf{q}$.

\section{Discussions}
In the above derivation, we get the superfluid density in terms of two-particle Green function (the Josephson relation) Eq.(\ref{spdensity}). In this section, we will show that for some cases, the above equation can be further simplified. In particular, if the system has only one gapless collective mode, the superfluid density can be give by the trace of two-particle Green function.
When the inversion symmetry is present, the superfluid density can be expressed in terms of the fluctuation matrix of superfluid order parameters.

\subsection{ only one gapless collective mode  }
When a system has unique gapless excitation near ground state, e.g., phonon, the Josephson relation can be generalized to multi-component system through a phase operator method as done in bosonic case \cite{zhangyicai2018}.
Here we know near the ground state, the phonon's excitation corresponds to total density oscillation. Due to the presence of superfluid order parameter, the density oscillation would couple phase oscillation of order parameter \cite{Ambegaokar1961}.  Furthermore, all the superfluid order parameters should share a common phase variation, i.e., $\delta \theta_\sigma(\textbf{r})=\delta\theta(\textbf{r})$. On the other hand, near the ground state, similarly as Eq.(\ref{deltaphi}), the fluctuation operator of superfluid order parameters may be expressed in terms of  phase operator $\hat{\theta}$ \cite{Lifshitz}
%\begin{align}
%\hat{\Delta}_{\sigma}(\textbf{r})=\Delta_{\sigma}e^{2i\hat{\theta}_{\sigma}(\textbf{r})},
%\end{align}
%where $\Delta_{\sigma}$ is the $\sigma-th$ superfluid order parameter in ground state.
%At long wave length limit, the variation of phase is small (the variations of amplitudes can be neglected \cite{Ambegaokar1961}), so we can get
\begin{align}
\delta\hat{\Delta}_{\sigma}(\textbf{r})\simeq 2i\Delta_\sigma\delta \hat{\theta}(\textbf{r}),
\label{basic22}
\end{align}
where $\Delta_{\sigma}$ is the $\sigma-th$ superfluid order parameter in ground state. Under perturbation $H'$ (see eq.(\ref{H1})), the variations of order  parameters can be obtained by averaging eq.(\ref{basic22}) with respect to the perturbed ground state.
Consequently, the variations for order parameters are $\delta\Delta_{\sigma}=i\Delta_{\sigma}\delta\theta(\textbf{r})$ with $\delta\theta(\textbf{r})=\langle\delta\hat{ \theta}(\textbf{r})\rangle$.

From the above equation, we get the fluctuation operators of order parameter in momentum space
\begin{align}
&\Delta_{\sigma,\textbf{q}}=2i\Delta_\sigma\hat{\theta}_{\textbf{q}},\notag\\
&\Delta^{\dag}_{\sigma,\textbf{q}}=-2i\Delta^{*}_{\sigma}\hat{\theta}^{\dag}_{\textbf{q}}=-2i\Delta^{*}_{\sigma}\hat{\theta}_{-\textbf{q}}.\notag\\
&\Delta_{\sigma,-\textbf{q}}=2i\Delta_\sigma\hat{\theta}_{-\textbf{q}},\notag\\
&\Delta^{\dag}_{\sigma,-\textbf{q}}=-2i\Delta^{*}_{\sigma}\hat{\theta}^{\dag}_{-\textbf{q}}=-2i\Delta^{*}_{\sigma}\hat{\theta}_{\textbf{q}},
\end{align}
where we use  $\hat{\theta}^{\dag}_{\textbf{q}}=\hat{\theta}_{-\textbf{q}}$ for real phase field $\theta(\textbf{r})$ ($\textbf{q}\neq0$) .
From definitions of the $G_{II}$ and $F_{II}$ in eq.(\ref{definition}), we get
\begin{align}
 &G_{II\sigma\sigma'}(\textbf{q},0)=-4Z\Delta_\sigma\Delta^{*}_{\sigma'},\notag\\ &F_{II\sigma'\sigma}(\textbf{q},0)=4Z\Delta_{\sigma'}\Delta_{\sigma} ,
\end{align}
where $Z\equiv\sum_n[\frac{|\langle0|\hat{\theta}_{\textbf{q}}|n\rangle|^2}{\omega_{n0}}+\frac{|\langle0|\hat{\theta}_{-\textbf{q}}|n\rangle|^2}{\omega_{n0}}]>0$ is a  real number.  When the number of superfluid order parameters is one and $\Delta$ is real, the relation $F_{II}(\textbf{q},0)=-G_{II}(\textbf{q},0)$ is obtained as $q\rightarrow0$ in usual two-component fermion superfluid.

From eqs.(\ref{basic}) and (\ref{basic2}), we get
\begin{align}
 &&&2i\delta\theta(\textbf{r})=-4Z\sum_\sigma [ \Delta^{*}_{\sigma} x_\sigma-\Delta_\sigma x^{*}_{\sigma}],\notag\\
 &&&=-4Z\sum_\sigma 2i\alpha_\sigma sin(\textbf{q}\cdot\textbf{r}+\phi_\sigma),\notag\\
 &&& \delta \textbf{j}(\textbf{r})=-\frac{2\textbf{q}}{q^2}\sum_{\sigma}[ \Delta^{*}_{\sigma} x_\sigma+\Delta_\sigma x^{*}_{\sigma}],\notag\\
 &&&=-\frac{2\textbf{q}}{q^2}\sum_{\sigma}[2\alpha_\sigma cos(\textbf{q}\cdot\textbf{r}+\phi_\sigma)]\notag\\
 &&&=\frac{\vec{\nabla} \delta\theta(\textbf{r})}{q^2Z}=\frac{\delta \textbf{v}_s}{q^2Z},
\end{align}
where we take $ \Delta^{*}_{\sigma} \xi_\sigma\equiv\alpha_\sigma e^{i\phi_\sigma}$ with amplitude $\alpha_{\sigma}$, phase $\phi_{\sigma}$.
So the superfluid density is
\begin{align}
\rho_s=\frac{1}{q^2Z}.
\end{align}

On the other hand, we know
\begin{align}
tr G_{II}(\textbf{q},0)\equiv\sum_\sigma G_{\sigma\sigma}(\textbf{q},0)=-4Z \sum_\sigma|\Delta_{\sigma}|^2.
\end{align}
 So finally we get the Josephson relation
\begin{align}
\rho_s=-{\rm lim_{q\rightarrow0}}\frac{4\sum_\sigma|\Delta_{\sigma}|^2}{q^2 trG_{II}(\textbf{q},0)},
\label{spdensity2}
\end{align}
where $G_{II}(\textbf{q},0)$ is two-particle normal Green function (matrix) at zero-frequency. When the number of order parameters $n=1$, the above equation is reduced to the Josephson relation for usual  two component fermion superfluid \cite{Taylor,Dawson}.

%The above formula for superfluid density can be also applied to multi-band system, for example, Haldane model \cite{normaldensity,zhangyicai}, dice lattice \cite{} or double-layered graphene conductor\cite{Lin1} .

\subsection{superfluid density and Pairing fluctuations matrix }
In this section, we would give a connection between the two-particle Green function and the pairing fluctuation matrix based on BCS mean field theory and Gaussian fluctuation approximation \cite{Taylor}.
 Further more, if the system has spatial inversion symmetry, the coefficient matrix $\textbf{G}_{\textbf{II}}$ is directly proportional to the inverse of pairing fluctuation matrix.

Fist of all, we discuss the results for usual two-component fermion superfluid.
In the following, we assume the pairing gap can be decomposed as the mean field value $\Delta$ and small fluctuation $ \delta\Delta(x)$ ($\Delta(x)=\Delta+\delta\Delta(x)$).
Next we consider the fluctuations about the mean-field results. Expanding action $S$ to second order of $\delta\Delta$ ,
the partition function \cite{Engelbrecht1997,Hu2006, Diener2008}
\begin{eqnarray}
 && Z\approx e^{-S_0}\int D \eta^{\dag}_q D\eta_q e^{-\delta S},
\end{eqnarray}
where $S_0$ is the mean-field contribution and Gaussian fluctuation part
\begin{eqnarray}
 && \delta S=\frac{1}{2}\sum_{\textbf{q},n}\eta^{\dag}_q M(q)\eta_q,\notag\\
 &&=\frac{1}{2}\sum_{\textbf{q},n}\eta^{\dag}_q\left(                 %×óÀšºÅ
 \begin{array}{cccc}   %žÃŸØÕóÒ»¹²3ÁÐ£¬Ã¿Ò»ÁÐ¶ŒŸÓÖÐ·ÅÖÃ
   M^o_{11}(q)  &  M^o_{12}(q)   \\  %µÚÒ»ÐÐÔªËØ
  M^o_{21}(q)    &      M^o_{22}(q)   \notag
\end{array}\right)\eta_q,\notag\\
&&=\sum_{\textbf{q}, n>0}\eta^{\dag}_q\left(                 %×óÀšºÅ
 \begin{array}{cccc}   %žÃŸØÕóÒ»¹²3ÁÐ£¬Ã¿Ò»ÁÐ¶ŒŸÓÖÐ·ÅÖÃ
   M^o_{11}(q)  &  M^o_{12}(q)   \\  %µÚÒ»ÐÐÔªËØ
  M^o_{21}(q)    &      M^o_{22}(q)   \notag
\end{array}\right)\eta_q\notag\\
\end{eqnarray}
with pairing fluctuation fields $\eta^{\dag}_q=[\Delta^{*}_q,\Delta_{-q}]$ and $q=(\textbf{q},i\omega_n)$. The
 fluctuation matrix $M$ is $2\times2$ matrix
\begin{eqnarray}
 &&M_{11}(\textbf{q},i\omega_n)=\frac{1}{\beta}\sum_{\textbf{k},n'}G_{11}(k+q)G_{22}(k)+\frac{1}{g},\notag\\
 &&M_{12}(\textbf{q},i\omega_n)=\frac{1}{\beta}\sum_{\textbf{k},n'}G_{12}(k+q)G_{12}(k),\notag\\
&& M_{21}(\textbf{q},i\omega_n)=M_{12}(\textbf{q},i\omega_n),\notag\\
 &&M_{22}(\textbf{q},i\omega_n)=M_{11}(-\textbf{q},-i\omega_n),
\end{eqnarray}
where $G_{ij}(k)$ is matrix element of Nambu-Gorkov Green function and $g$ is the interaction strength between particles. $\omega_n=2n\pi/\beta$ ($n\in Z$) is Matsubara frequency, $\beta=1/T$ is inverse temperature and $k=(\textbf{k},i\omega_{n'})$.
 The collective modes are given by zeros of determinant $Det|M(\textbf{q},i\omega_n\rightarrow \omega+i0^+)|=0$. As $q\rightarrow0$, the collective mode is the Anderson-Bogoliubov phonon, which characterizes the density oscillations of superfluid.
With the action $\delta S$ (Gaussian weight), the correlation function in imaginary time (average values of quadratic terms) can be calculated \cite{yicai2017}, i.e.,
 \begin{eqnarray}
 &&\langle \Delta^{*}_q \Delta_{q}\rangle=(M^{-1})_{11},\notag\\
 &&\langle \Delta_{-q} \Delta_q \rangle=(M^{-1})_{12},\notag\\
&& \langle \Delta^{*}_{q}\Delta^{*}_{-q} \rangle=(M^{-1})_{21},\notag\\
 &&\langle  \Delta_{-q}\Delta^{*}_{-q}\rangle=(M^{-1})_{22}.
\end{eqnarray}
 On the other hand, we know that the above correlation function in imaginary time has one extra minus sign comparing with Green function.
So the two-particle Green functions can be obtained
 \begin{eqnarray}
 &&-G_{II}(q)=\langle \Delta^{*}_q \Delta_{q}\rangle=(M^{-1})_{11},\notag\\
 &&-F_{II}(q)=\langle \Delta_{-q} \Delta_q \rangle=(M^{-1})_{12},\notag\\
 &&-F^{*}_{II}(q)=\langle \Delta^{*}_{q}\Delta^{*}_{-q} \rangle=(M^{-1})_{21},\notag\\
 &&-G_{II}(-q)=\langle  \Delta_{-q}\Delta^{*}_{-q}\rangle=(M^{-1})_{22}.
\end{eqnarray}
In addition, if the system has inversion symmetry, i.e.,
 \begin{eqnarray}
 &&G_{II}(-\textbf{q},i\omega_n)=G_{II}(\textbf{q},i\omega_n),\notag\\
 &&F_{II}(-\textbf{q},i\omega_n)=F_{II}(\textbf{q},\omega_n),
\end{eqnarray}
further more, according to Eq.(\ref{definition}), when $\omega\pm i\epsilon=0$, the Green function satisfy
\begin{align}
 &G^{*}_{II\sigma\sigma'}(\textbf{q},0)=G_{II\sigma'\sigma}(\textbf{q},0),\notag\\
 &F_{II\sigma\sigma'}(\textbf{q},0)= F_{II\sigma'\sigma}(-\textbf{q},0),
\end{align}
then the coefficient matrix of Green function $\textbf{G}_{\textbf{II}}$ can be expressed in terms of the inverse of $M$
  \begin{eqnarray}
 &&\textbf{G}_{\textbf{II}}=\left( \!\!\!                %×óÀšºÅ
 \begin{array}{cccc}   %žÃŸØÕóÒ»¹²3ÁÐ£¬Ã¿Ò»ÁÐ¶ŒŸÓÖÐ·ÅÖÃ
  G_{II}(q)  & F_{II}(q) \\  %µÚÒ»ÐÐÔªËØ
   F^{*}_{II}(q)&  G^{*}_{II}(q)\\  %µÚ¶þÐÐÔªËØ
\end{array}\!\!\!\right),\notag\\
&&=\left( \!\!\!                %×óÀšºÅ
 \begin{array}{cccc}   %žÃŸØÕóÒ»¹²3ÁÐ£¬Ã¿Ò»ÁÐ¶ŒŸÓÖÐ·ÅÖÃ
  G_{II}(q)  & F_{II}(q) \\  %µÚÒ»ÐÐÔªËØ
   F^{*}_{II}(q)&  G_{II}(-q)\\  %µÚ¶þÐÐÔªËØ
\end{array}\!\!\!\right)=-M^{-1}.
\end{eqnarray}
 So the superfluid density is given by
 \begin{eqnarray}
 \rho_s&&={\rm lim_{q\rightarrow0}}\frac{-4}{q^2}(\Delta^{t*},\Delta^t).\!\!\left( \!\!\!                %×óÀšºÅ
 \begin{array}{cccc}   %žÃŸØÕóÒ»¹²3ÁÐ£¬Ã¿Ò»ÁÐ¶ŒŸÓÖÐ·ÅÖÃ
  I  & 0 \\  %µÚÒ»ÐÐÔªËØ
   0&  -I\\  %µÚ¶þÐÐÔªËØ
\end{array}\!\!\!\right).\textbf{G}_{\textbf{II}}^{-1}.\!\!\left( \!\!\!                %×óÀšºÅ
 \begin{array}{cccc}   %žÃŸØÕóÒ»¹²3ÁÐ£¬Ã¿Ò»ÁÐ¶ŒŸÓÖÐ·ÅÖÃ
  I  & 0 \\  %µÚÒ»ÐÐÔªËØ
   0&  -I\\  %µÚ¶þÐÐÔªËØ
\end{array}\!\!\!\right)\!\!.\!\!\left(\!\!
 \begin{array}{cccc}   %žÃŸØÕóÒ»¹²3ÁÐ£¬Ã¿Ò»ÁÐ¶ŒŸÓÖÐ·ÅÖÃ
   \Delta \\  %µÚÒ»ÐÐÔªËØ
   \Delta^*\\  %µÚ¶þÐÐÔªËØ
\end{array}\!\!\right),\notag\\
&&={\rm lim_{q\rightarrow0}}\frac{4}{q^2}(\Delta^{t*},\Delta^t).\!\!\left( \!\!\!                %×óÀšºÅ
 \begin{array}{cccc}   %žÃŸØÕóÒ»¹²3ÁÐ£¬Ã¿Ò»ÁÐ¶ŒŸÓÖÐ·ÅÖÃ
  I  & 0 \\  %µÚÒ»ÐÐÔªËØ
   0&  -I\\  %µÚ¶þÐÐÔªËØ
\end{array}\!\!\!\right).M(q).\!\!\left( \!\!\!                %×óÀšºÅ
 \begin{array}{cccc}   %žÃŸØÕóÒ»¹²3ÁÐ£¬Ã¿Ò»ÁÐ¶ŒŸÓÖÐ·ÅÖÃ
  I  & 0 \\  %µÚÒ»ÐÐÔªËØ
   0&  -I\\  %µÚ¶þÐÐÔªËØ
\end{array}\!\!\!\right)\!\!.\!\!\left(\!\!
 \begin{array}{cccc}   %žÃŸØÕóÒ»¹²3ÁÐ£¬Ã¿Ò»ÁÐ¶ŒŸÓÖÐ·ÅÖÃ
   \Delta \\  %µÚÒ»ÐÐÔªËØ
   \Delta^*\\  %µÚ¶þÐÐÔªËØ
\end{array}\!\!\right).\notag\\
\end{eqnarray}

Next, for the case of several superfluid order parameters, the proof is similar.
Assuming the number of superfluid order parameters are arbitrary $m$, then Gaussian fluctuation part
\begin{eqnarray}
 \delta S=\sum_{\textbf{q},n>0}\eta^{\dag}_q M(q)\eta_q,
% &&=\frac{1}{2}\sum_{\textbf{q},n}\eta^{\dag}_q\left(                 %×óÀšºÅ
% \begin{array}{cccc}   %žÃŸØÕóÒ»¹²3ÁÐ£¬Ã¿Ò»ÁÐ¶ŒŸÓÖÐ·ÅÖÃ
%   M_{11}(q)  &  M_{12}(q) & M_{13}(q)  &  M_{14}(q)  \notag \\  %µÚÒ»ÐÐÔªËØ
%  M_{21}(q)    &      M_{22}(q) & M_{23}(q)    &      M_{24}(q)  \notag\\
%   M_{31}(q)  &  M_{32}(q) & M_{33}(q)  &  M_{34}(q)  \notag \\
%  M_{41}(q)    &      M_{42}(q) & M_{43}(q)    &      M_{44}(q)  \notag
%\end{array}\right)\eta_q
\end{eqnarray}
where $M(q)$ is a $2m\times2m$ fluctuation matrix, and pairing fluctuation fields $\eta^{\dag}_q=[\Delta^{*}_{1q},\Delta^{*}_{2q},...,\Delta^{*}_{mq},\Delta_{1,-q},\Delta_{2,-q},...,\Delta_{m,-q}]$ .
With the action $\delta S$ (Gaussian weight), the  correlation functions (and the matrix elements of Green function) in imaginary time can be calculated, i.e.,
 \begin{eqnarray}
 &&-G_{II,ij}(q)=\langle \Delta^{*}_{jq} \Delta_{iq}\rangle=(M^{-1})_{i,j},\notag\\
 &&-F_{II,ij}(q)=\langle \Delta_{j,-q} \Delta_{iq} \rangle=(M^{-1})_{i,m+j},\notag\\
&& -F^{*}_{II,ji}(q)=\langle \Delta^{*}_{jq}\Delta^{*}_{i,-q} \rangle=(M^{-1})_{m+i,m+j},\notag\\
 && -G_{II,ji}(-q)=\langle  \Delta_{j,-q}\Delta^{*}_{i,-q}\rangle=(M^{-1})_{m+i,m+j}.
\end{eqnarray}
Further more, in the presence of inverse symmetry,
 the Green function coefficient matrix can be obtained through the fluctuation matrix, i.e., $\textbf{G}_{\textbf{II}}=-M^{-1}$.
 So the superfluid density is given by
 \begin{eqnarray}
 \rho_s\!\!=\!\!{\rm lim_{q\rightarrow0}}\frac{4}{q^2}(\Delta^{t*},\Delta^t).\!\!\left( \!\!\!                %×óÀšºÅ
 \begin{array}{cccc}   %žÃŸØÕóÒ»¹²3ÁÐ£¬Ã¿Ò»ÁÐ¶ŒŸÓÖÐ·ÅÖÃ
  I  & 0 \\  %µÚÒ»ÐÐÔªËØ
   0&  -I\\  %µÚ¶þÐÐÔªËØ
\end{array}\!\!\!\right).M(q).\!\!\left( \!\!\!                %×óÀšºÅ
 \begin{array}{cccc}   %žÃŸØÕóÒ»¹²3ÁÐ£¬Ã¿Ò»ÁÐ¶ŒŸÓÖÐ·ÅÖÃ
  I  & 0 \\  %µÚÒ»ÐÐÔªËØ
   0&  -I\\  %µÚ¶þÐÐÔªËØ
\end{array}\!\!\!\right)\!\!.\!\!\left(\!\!
 \begin{array}{cccc}   %žÃŸØÕóÒ»¹²3ÁÐ£¬Ã¿Ò»ÁÐ¶ŒŸÓÖÐ·ÅÖÃ
   \Delta \\  %µÚÒ»ÐÐÔªËØ
   \Delta^*\\  %µÚ¶þÐÐÔªËØ
\end{array}\!\!\right).
\label{spdensity3}
\end{eqnarray}

\section{An Example: Superfluid density in Haldane-Hubbard model }
As an application of the Josephon relation, e.g., Eqs.(\ref{spdensity2}) and (\ref{spdensity3}), the superfluid density is calculated for Haldane-Hubbard model in two-component Fermi gas with on-site attractive interaction $-U$ ($U>0$) \cite{zhangyicai2017}. The Hamiltonian for Haldane-Hubbard model is
 \begin{eqnarray}
 H=\sum_{ij\sigma}t_{ij}c_{i}^{\dag}c_j+\sum_{i\sigma}(M\epsilon_i-\mu)n_{i\sigma}-U\sum_in_{i\uparrow}n_{i\downarrow},
\end{eqnarray}
 where $\mu$ is chemical potential, and $n_{i\sigma}=c^{\dag}_{i\sigma}c_{i\sigma}$ is particle number operator.  $\epsilon_i=1(-1)$ for sublattice A (B) and $M$ is energy offset between sublattice A and B. $t_{ij}$ is hopping amplitude between lattice sites $i$ and $j$, which is $t$ for nearest neighbor sites, $t'e^{-i\phi}(t'e^{i\phi})$ for clockwise (anticlockwise) hopping between next-nearest neighbor sites \cite{zhangyicai2017}. The distance between nearest neighbor sites is $a$.

 In the following, we focus on the case of $\phi=\pi/2$ and $M=0$, where the inversion symmetry is not broken . When the filling factor is half-filling ($n=2$ particles per unit cell), for weakly interacting case, the system is a Chern insulator, and the superfluid order parameters vanishes. Only when the interaction is strong enough, the system enters into a superfluid phase and superfluid order parameter $\Delta\neq0$. Away from half-filling, the system is usually in a superfluid phase \cite{zhangyicai2017}.
 In addition, it is found that, there exist only one gapless excitation near $q=0$, which corresponds to total density oscillations. The superfluid density can be also calculated with current-current correlation \cite{Baym,zhangyicai2016,Hazra2019}  or  phase twist method \cite{zhangyicai2020,Liang2017}.  Assuming the superfluid order parameters undergo a phase variation, e.g, $\triangle_i\rightarrow \Delta_ie^{2i\textbf{q}\cdot\textbf{r}_i}$,
The superfluid density (particle number per unit cell) tensor $\rho_{sij}$ can be written as
 \begin{eqnarray}
 \rho_{sij}= \frac{\partial^2 \Omega(\textbf{q})}{\partial q_i \partial q_i}|_{q\rightarrow0}.
 \label{phasetwist}
\end{eqnarray}
where $\Omega$ is thermodynamical potential (per unit cell) in grand canonical ensemble.

%In BCS mean field theory, the Hamiltonian is decoupled for different momentum $k$. The above variation of order parameters can be transferred to the single-particle Hamiltonian by a gauge transformation, $\psi\rightarrow \psi e^{i\textbf{q}\cdot\textbf{r}}$
%Then the Hamiltonian is $\textbf{q}-$ dependent, i.e., $H\rightarrow H(\textbf{q})$,
%When temperature is zero, the above superfluid density can be calculated through perturbation theory
% \begin{eqnarray}
% \rho_{sij}= \sum_{\textbf{k}}\langle 0|\frac{\partial^2H(\textbf{k})}{\partial k_i\partial k_j}|0\rangle
% +\sum_n\frac{\langle0|\frac{\partial  H(\textbf{k})}{\partial k_i}|n\rangle\langle n|\frac{\partial  H(\textbf{k})}{\partial k_i}|0\rangle}{(E_0-E_n)^2}
% \end{eqnarray}
%

\begin{figure}
\begin{center}
\includegraphics[width=1.0\columnwidth]{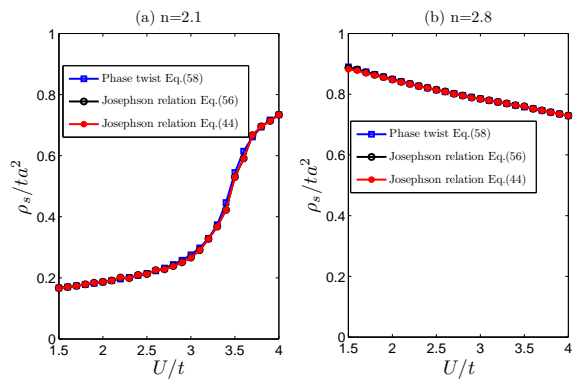}
\end{center}
\caption{  Superfluid densities (particle number per unit cell) are plotted as functions of interaction $U$  (a) filling factor $n=2.1$, $t'=0.15t$, $M=0$ and $\phi=\pi/2$; (b) filling factor $n=2.8$, $t'=0.15t$, $M=0$ and $\phi=\pi/2$. The the superfluid density in the three curves are obtained through three different formulas, i.e., Eqs.(\ref{phasetwist}), (\ref{spdensity3}) and (\ref{spdensity2}).  }
\label{pdm0}
\end{figure}
\begin{figure}
\begin{center}
\includegraphics[width=1.0\columnwidth]{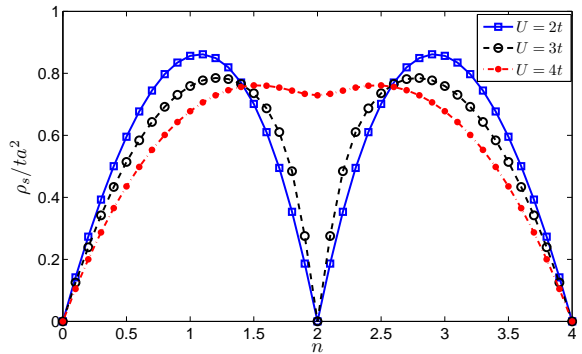}
\end{center}
\caption{  Superfluid densities (particle number per unit cell) are plotted as functions of the filling factor $n$ (particle number per unit cell)  with $t'=0.15t$, $M=0$ and $\phi=\pi/2$. The three curves correspond to interaction $U=2t,3t$ and 4t, respectively.  }
\label{pdm0}
\end{figure}

Fig.1 shows the evolutions of superfluid density as the interaction increases. The superfluid density is obtained with three different formulas, i.e. Eqs.(\ref{phasetwist}), (\ref{spdensity3}) and (\ref{spdensity2}). First of all, it is found that the superfluid density is isotropic and behaves as a scalar in two-dimensional space, i.e., $\rho_{sij}=diag(\rho_s,\rho_s)$, which  is a consequence of $C_3$ rotational symmetry of honeycomb lattice \cite{Liang2017}.   Secondly, the results from the Eq.(\ref{spdensity3}) and (\ref{spdensity2}) are consistent with the results from phase twist method Eq.(\ref{phasetwist}). In addition, When filling factor $n=2.1$, the superfluid density increases as the interaction get strong. However when the filling factor $n=2.8$, the superfluid density get smaller and smaller when the interaction increases.  Such a interesting feature is also reflected in Fig.2, that when the filling factor is near half-filling ($n=2$), the superfluid density increases as the interaction get strong. However, when the filling is far away from half-filling, the situation is reversed (see Fig.2).

Fig.2 shows the superfluid density as functions of filling factor ($n=0\rightarrow 4$). From the Fig.2, we can see that the superfluid density is symmetric with respect to half-filling ($n=2$) due to particle-hole symmetry \cite{zhangyicai2017}. For weak interaction cases ($U=2t$ or $U= 3t$) and half filling, the system is a insulator (see Ref.\cite{zhangyicai2017}), the superfluid density vanishes (see Fig.2). For fully occupied case ($n=4$ particle per unit cell), the system is equivalent to the fully empty case  ($n=0$) due to the particle-hole symmetry, the system is also a insulator. Consequently, superfluid density is also zero. When the filling factor falls into in the middle ($n\approx 3$), the superfluid density reaches its maximum value.
The appearance of double dome structure for weak interactions in Fig.2 is in qualitative agreement with the results obtained through dynamical mean-field theory (DMFT) in Ref. \cite{Liang2017}.

\section{summary}
In conclusion, we investigate the Josephson relation for  a general multi-component fermion superfluid.
 It is found that the superfluid density is given in terms of two-particle Green functions. When the superfluid has only one gapless collective excitation, the Josephson relation can be simplified, which is given  in terms superfluid order parameters and trace of Green function.
Within BCS mean field theory and Gaussian fluctuation approximation, the matrix elements of Green function can be given in terms of pairing fluctuation matrix elements.
Further more, in the presence of inversion symmetry, it is shown that the two-particle Green function is directly proportional the inverse of pairing fluctuation matrix. The formulas for superfluid density is quite general, which can be also applied to multi-band superfluid with complex spectra in lattice.

Josephson relation for multi-component fermion superfluid  provides a general method for calculations on superfluid densities in terms of two-particle Green functions and fluctuation matrix. Our work would be useful for investigations on  the superfluid properties of multi-component (or multi-band) superfluid system with complex pairing structures.

\acknowledgements
%We thank Shizhong Zhang for useful discussions.
 This work was supported by the NSFC under Grants
No.11874127.
We also acknowledge the supports of startup grant from Guangzhou University.
%\appendix*

 %\section{Two-particle Green function and the fluctuation matrix }

\appendix*

\end{document}